\documentclass[a4paper,footinbib]{revtex4}

\usepackage{amssymb}
\usepackage{amsmath}
\usepackage{amsbsy}
\usepackage{graphicx}

\begin{document}
\title{Effect of an electric field on a Leidenfrost droplet}

\author{Franck Celestini}
   \email{Franck.Celestini@unice.fr}
   \affiliation{Laboratoire de Physique de la Mati\`ere Condens\'ee,
  UMR 7366, CNRS, Universit\'e de Nice Sophia-Antipolis,
   Parc Valrose  06108 Nice Cedex 2, France}

\author{Geoffroy Kirstetter}
   \affiliation{Laboratoire de Physique de la Mati\`ere Condens\'ee,
  UMR 7366, CNRS, Universit\'e de Nice Sophia-Antipolis,
   Parc Valrose  06108 Nice Cedex 2, France}

\date{\today}

\begin{abstract}
We experimentally investigate the effect of an electric field applied between a Leidenfrost droplet and the heated  substrate on which it is levitating. We quantify the electro-Leidenfrost effect by imaging the interference fringes between the liquid-vapour and vapour-substrate interfaces. The increase of the voltage induces a decrease of the vapour layer thickness. Above a certain critical voltage the Leidenfrost effect is suppressed and the drop starts boiling. Our study characterizes this way to control and/or to avoid the Leidenfrost effect that is undesirable in many domains such as metallurgy or nuclear reactor safety.
\end{abstract}

\maketitle

The first observation of the Leidenfrost  effect has been reported  more  than two centuries ago \cite{leiden} by the German scientist J. G. Leidenfrost. He showed that a drop  deposited on a sufficiently hot plate  keeps its spheroidal shape and levitates on its own vapour. Because of the weak heat conduction of the vapour as compared to the one of the liquid, the evaporation time is considerably increased and reaches a maximum at $T_f$, the so-called Leidenfrost temperature. For example, a millimetric water drop has a lifetime of several minutes on a substrate kept at $T = 250 ^\circ C$ while it evaporates in less than one second when deposited on the same substrate but kept at $150  ^\circ C$. This subject 
 is still motivating numerous researches. From a fundamental point of view,
studies are essentially motivated by the non-wetting situation, considered  as the limit of a perfect super-hydrophobicity, and by the high mobility of the droplet due to
the weak friction with the substrate on which it is levitating \cite{quere,linke,toro,clanet}. For many  industrial applications the Leidenfrost effect is rather undesirable. As stressed above,  the vapour layer in between the liquid and the substrate dramatically reduces the heat conduction. This has important consequences for example in metallurgy for controlling the quenching process of alloys \cite{bernardin} and in the safety of nuclear reactors \cite{vandam}.

In this letter, we study the effect of a tension applied between a Leidenfrost drop and the substrate on which it is levitating. The vapour layer thickness decreases
with the applied voltage and a rather low critical voltage (typically $40 V$ for a millimetric drop)  makes it possible to suppress the Leidenfrost effect.  In spite of numerous studies on the electro-hydrodynamic enhancement of heat transfer 
in fluid flow \cite{allen}, only one study  \cite{patent} has considered this possibility but quantitative studies have not been performed. In this communication,  first we present the experimental set-up consisting in the observation of the interferences fringes between the liquid and the substrate. This method  allows a precise measurement of the vapour layer thickness. Without any applied voltage we put in evidence an asymmetric underneath profile of the Leidenfrost drop. We then present our  results on the relative variation of the vapour layer thickness sustaining the Leidenfrost droplet as a function of the applied voltage.  A model is proposed and predicts that the vapour thickness decreases linearly with the square of the applied voltage. This  model  successfully reproduces  the experimental measures  performed on water droplets. We finally draw some conclusions on the electo-Leidenfrost phenomenon presented in this letter.

\begin{figure}
  \includegraphics[width=.45\textwidth]{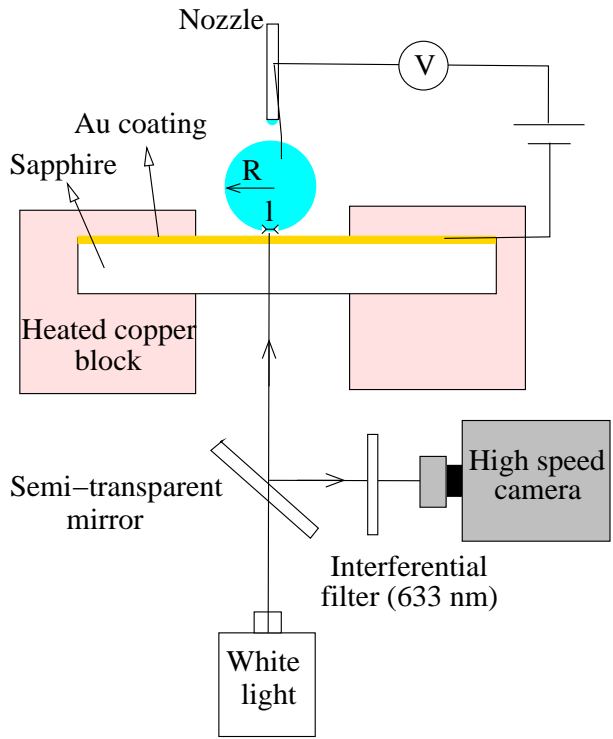}
  \caption{Experimental set-up.}
  \label{fig_set-up}
\end{figure}

The experimental set-up is presented in Fig. \ref{fig_set-up}. 
A copper block  is kept at a controlled temperature $T$. A sapphire substrate  is inserted in the copper block.
  To ensure its electrical conductivity, the substrate is coated with a thin (60 nm width) gold film.  A water droplet of radius $R$ is generated from a nozzle  where a thin tungsten wire ($150 \mu m$ radius) is fixed in order to ensure the electrical conduction with the drop. A low frequency  ($0.5$ Hz) voltage (amplitude ranging between $0$ and $30 V$) is applied  between the drop  and the upper coated
    part of the sapphire. The sapphire is illuminated by the bottom with a white light source. A  semi-transparent mirror together with
     an interferometer filter ($\lambda = 633$ nm, width $=30 nm$) are used to visualized the interference fringes between the liquid-vapour and vapour-sapphire interfaces.    
      A high speed camera is used to record
      the spatial and time  evolution of the interference fringes. The height difference between two white (or black) fringes  is of $\lambda/2 = 0.317 \mu m$.  During the redaction of this letter we have been aware that a similar experimental set-up was used to study the fluctuation
       of the vapour thickness of a Leidenfrost drop \cite{preprint}. 
      
       The interference fringes are present in the nearest contact zone of the droplet with the substrate where the vapour film is thin enough. We assume that the drop has the shape of a sphere flattened on its bottom. The disc between the drop and the film vapour (the so-called Laplacian disc) has a radius $l$ related to the radius $R$ of the drop through \cite{yves} :
 
 \begin{equation}
 l \propto R^2 / \kappa^{-1}
 \label{disc}
 \end{equation}
 
 where $\kappa^{-1} = \sqrt{\gamma/\rho_l g}$ is the capillary length, $\gamma$ the surface tension, $g$ the acceleration of gravity and $\rho_l$ the mass liquid density. This relation is obtained assuming $R\ll\kappa^{-1}$ and by balancing the gravity force with the Laplace pressure within the drop.

\begin{figure}
  \includegraphics[width=.5\textwidth]{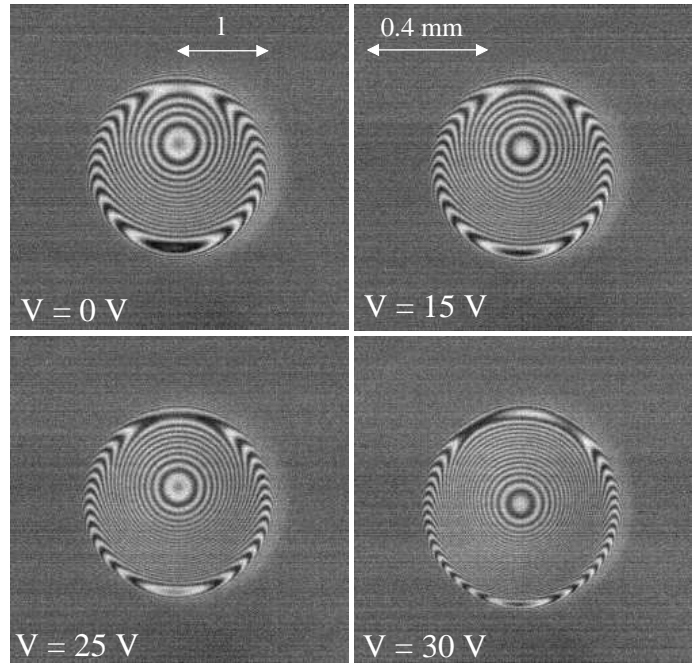}
  \caption{ Interferences patterns at the bottom of a Leidenfrost droplet of radius $R\simeq 1$ mm and Laplacian disk radius $l$.
  The interference patterns are represented for $V=0$, $15$, $25$ and $30 V$. }
  \label{fig_photo}
\end{figure}

We present in Fig. \ref{fig_photo} the interference patterns observed at the bottom of a Leidenfrost droplet of radius $R\simeq 1 mm$, contact disc radius $l\simeq 0.3 mm$ and a temperature sapphire substrate kept at $280 ^\circ C$. The  pictures are taken for $V=0$, $15$, $25$ and $30 V$. We can observe that the fringes are not centred on the drop center. The height profile of the vapour pocket below the drop is therefore  not axis-symmetric.  Previous studies \cite{duchemin,eggers,charpin} have both theoretically and numerically investigated the equations governing the vapour thickness underneath a Leidenfost droplet. Nevertheless, all studies have supposed an axis-symmetric profile. The asymmetric shape has been systematically observed and is more important when the radius of the droplet decreases. As can be observed on Fig. \ref{fig_photo}, the electric field tends to reduce the asymmetry and
to increase the value of the Laplacian disk radius. 
 
\begin{figure}
  \includegraphics[width=.6\textwidth]{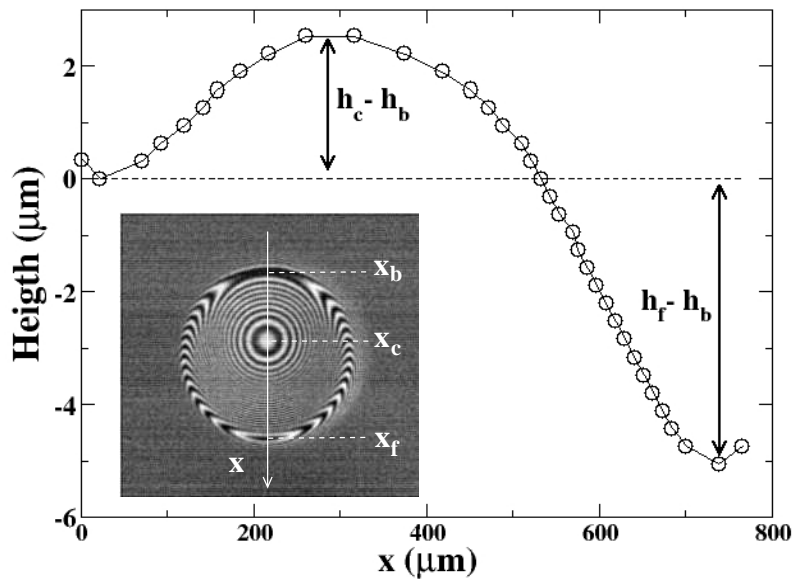}
  \caption{Height profile along  the $x$ diameter axis  (as sketched in inset). The height of the Leidenfrost droplet is minimum on its front ($x=x_f$), maximum on its center ($x=x_c$) and has a local minimum on its back ($x=x_b$). The present profile is obtained without any applied voltage and for a substrate
   temperature kept at $T=280 ^\circ C$. }
  \label{fig_prof}
\end{figure}

As sketched in the inset of Fig. \ref{fig_prof} we define an $x$ diameter axis. At the back of the drop ($x=x_b$) the height presents a local minimum $h=h_b$, on its the center ($x=x_c$)  the height is maximum at $h=h_c$ and  on the front of the droplet ($x=x_c$) the height reaches a minimum at $h=h_f$. The height profile along the $x$
  axis is measured and represented in  Fig. \ref{fig_prof} for a droplet at $V=0$. We take  $h_b=h(x_b=0)=0$ as the
   reference for our relative height interferential measurements. The height difference between the center and the front is  $  \delta h_{cf}\simeq 8 \mu m $
    while the one between the center and the back  $ \delta h_{cb}\simeq 2.5 \mu m $. The asymmetry discussed above can  be estimated by the the
ratio $ \delta h_{cb}/ \delta h_{cf} \simeq 3.2$ and also by $(x_c-x_b)/(x_f-x_c) \simeq 1.5$. Both values are significantly different from the one expected for the axis-symmetric case considered in the previous theoretical studies cited just above. 

 To evaluate the effect of the applied voltage on the height profile we apply a modulated tension amplitude of the form :
 
 \begin{equation}
  V(t)= \frac{V_{max}}{2} (1+ cos(2\pi f (t-t_0))) \mathrm{.}
  \label{voltage}
  \end{equation}
  
\begin{figure}
  \includegraphics[width=.45\textwidth]{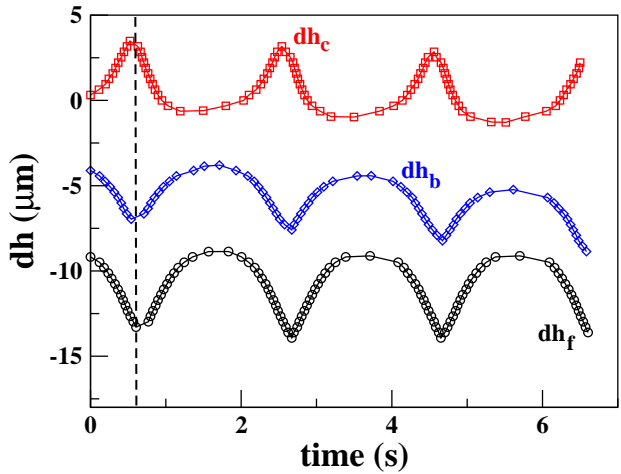}
  \caption{Time dependence of the relative heights, $dh_c$ (red squares), $dh_b$ (blue diamonds) and
   $dh_f$ (black circles) for a Leidenfrost droplet deposited on a substrate at $T=280 ^\circ C$. The dotted line indicates the time $t_0$ at which the applied voltage is maximum. The height reference is taken for $h_c(t=0)=0$, the applied voltage has an amplitude of $30 V$ and a frequency of $0.5$ Hz.}
  \label{fig_evot}
\end{figure}

   A low frequency modulation is used $f=0.5$ Hz and a maximum voltage $V_{max}=30 V$. We recorded the evolution
 of the interferences fringes and present in the supplementary material a movie recorded during several periods of the applied voltage. Using an image analysis, we compute the time evolution of the relative heights $h_b$,$h_c$ and $h_f$. The reference is taken for 
 $h_c(t=0)=0$. We represent the relative heights as a function of time in Fig. \ref{fig_evot}. The dotted line indicates the time $t_0$ at which the applied voltage is maximum. At $t_0$ both $h_f$ and $h_b$ reach a minimum value. The applied voltage induces a capacitive force that attracts the front and the bottom of the droplet toward the substrate. More surprisingly the center of the drop reaches a maximum height when the applied voltage is maximum. We qualitatively explain it considering that the decrease of $h_b$ and $h_f$  tends to increase the viscous pressure underneath the drop and therefore to increase height on its center, $h_c$.

It is out the scope of this paper to propose a model describing the evolution of the overall droplet profile under the action of the applied voltage. We limit our analysis considering a droplet flattened at its bottom on a disc of radius $l$ and a vapour width $h$. The relation between
$R$ and $l$ without any applied voltage is given by the equation (\ref{disc}).
 For a Leidenfrost droplet of radius $R$ the vapour thickness (without any applied voltage) $h_0$ can be estimated balancing the gravity force  with the viscous pressure
 of the  Poiseuille vapour flow between the drop and the substrate :

$$  F_p \propto \frac{\eta \lambda \Delta T l^4}{L \rho_v h^4} \mathrm{,}$$
   
   $\rho_v$ being the mass vapour density. It gives an expression for the equilibrium width $h_0$ satisfying :
   
 \begin{equation}
 \rho g R^3 = \frac{\eta \lambda \Delta T l^4}{L \rho_v h_0^4}
 \label{eqeq}
 \end{equation}
 
 In the previous expression,  $L$ is the latent heat, $\eta$ the vapour viscosity, $\lambda$ the thermal conductivity and $\Delta T$ the temperature
  difference between the substrate kept at temperature $T$ and the water drop at temperature $T_v=100 ^\circ C$. 
  It is important to note that we used the lubrication approximation to obtain the equilibrium thickness $h_0$. As can be seen in Fig. \ref{fig_prof} and as discussed recently \cite{eggers} there exist regions with high curvatures below the drop where the lubrication could not be  fully justified.
  
The force, $F_v$ due to the applied voltage  between the liquid-vapour and vapour-substrate interface can be approximated as the one in between the two armatures of a capacitor. It reads :

$$ F_v \propto - V^2 \epsilon l^2/h^2 $$
 
where $\epsilon$ is the dielectric constant. We consider the case for which the applied voltage induces a small variation $dh$ of the vapour thickness compared to its equilibrium value $h_0$. Using equation (\ref{eqeq}), the balance between the three different forces therefore leads to :

 \begin{equation}
dh \propto -  \frac{\epsilon V^2}{4} \frac{ \rho_v L}{ \eta \lambda \Delta T l^2}  h_0^3  \mathrm{•}
\label{eqmod}
 \end{equation}
 
It is worth mentioning  that, as for the electro-wetting effect \cite{mugele}, the deviation from the equilibrium situation for the electro-Leidenfrost effect is proportional to the square of the applied voltage.

 \begin{figure}
  \includegraphics[width=.45\textwidth]{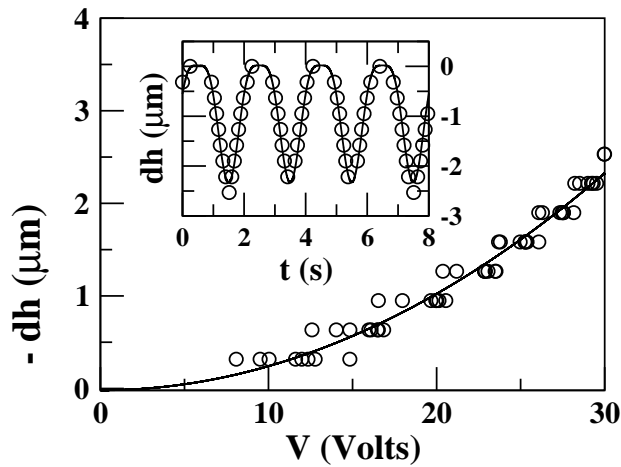}
  \caption{ Thickness variation $dh$ as a function of the applied voltage $V$ for a Leidenfrost droplet standing on a substrate kept at
 $280 ^\circ C$. The full line corresponds to the  best fit to the equation (\ref{eqmod}). Inset : relative variation of the thickness as a function of time for an applied voltage modulated in amplitude with a frequency $f=0.5$ Hz. The black line is the best fit to the equation (\ref{eqmod})}
  \label{fig_compa}  
\end{figure}

 We use  equation (\ref{voltage}) to eliminate  time in the data presented in Fig. \ref{fig_evot}. We can therefore plot the thickness variation $dh$ as a function of the applied voltage $V$. This is done in Fig. \ref{fig_evot} for the water droplet standing on a substrate kept at
  $280 ^\circ C$. In order to compare the experimental data with our simple model we assume in what follows that the mean thickness variation $dh \propto dh_f$. The full black line is a best fit to equation (\ref{eqmod}) with a single free parameter $h_0$, the equilibrium thickness. The best agreement is found for a value $h_0=18 \mu m$ which is close to previous measurements of the vapour thickness for millimetric drops \cite{quere}. In the inset of Fig. \ref{fig_evot}, we plot $dh$ as a function of time and the best fit to our model.  Dividing $V_{max}$ by $h_0$ gives an estimate of the maximum applied electric field $E_{max} \simeq 1.6$ $10^{6}$ $V m^{-1}$. This value is below the disruptive electric field and therefore justifies that we did not observe any electrical 
discharges during our experiments.  For the data obtained on substrates kept at lower temperatures we observed that the model underestimates $dh$*. At lower temperatures the equilibrium thickness is thinner and the condition  $dh \ll h_0$  is no longer valid for large values of the applied voltage. 
 
\begin{figure}
  \includegraphics[width=.55\textwidth]{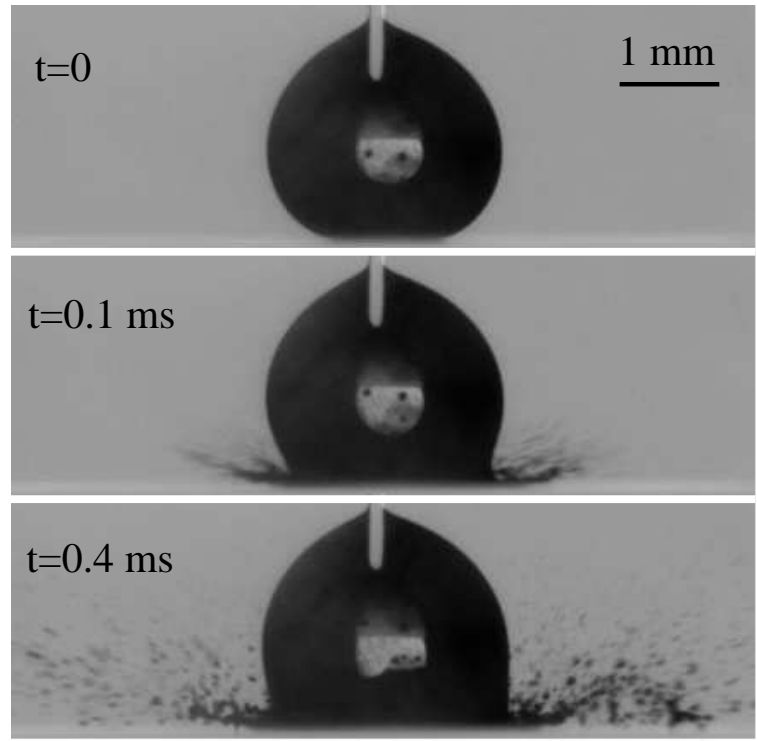}
  \caption{  A voltage of $40 V$ is suddenly applied to a Leidenfrost droplet at $t=0$. The three successive pictures display the boiling crisis of the droplet due to the applied electric field. }
  \label{fig_boiling}  
\end{figure}

  Applying a larger voltage suppress the Leidenfrost effect. For the millimetric drops considered in this study, an applied voltage of $40 V$ is sufficient to decrease to $0$ the value
   of $h_f$ and to let the drop starts boiling. We illustrate this effect by suddenly applying a voltage of $40 V$ on a millimetric droplet initially standing in the Leidenfrost state on a copper substrate. We represent in Fig. \ref{fig_boiling} three successive pictures taken at $t=0$, $0.1$ and $0.4 ms$. The voltage is applied at $t=0$ and we can observe the boiling crisis of the droplet. The movie M2 in the supplementary material displays the evolution of the droplet recorded at $90 000$ frame per second.  The critical voltage for suppressing the Leidenfrost effect could be simply evaluated combining equations (\ref{eqeq}) and (\ref{disc}) with the condition $dh/h_0 = 1$. Nevertheless this would underestimate the critical voltage. First,  equation (\ref{eqeq}) 
    is obtained assuming $dh/h_0 \ll 1$ and, as can be seen in \ref{fig_photo} and in the movie M1 of the supplementary material, the applied voltage also tends to increase the value of $l$. This effect has not be taken into account in our simple model. 
  
  In this communication we have investigated the effect of an electric field applied between a Leidenfrost drop and the substrate on which it is standing. Using an interferential technique we have measured the thickness of the vapour layer as a function of the applied tension. The experimental data are in good agreement with the simple model proposed that predicts a decrease of the vapour layer thickness proportional to the square of the applied voltage.
  For a millimetric drop, applying a voltage of order $40 V$ permits to suppress the Leidenfrost effect and to let the drop starts boiling. This study should find applications in domains where the Leidenfrost effect has to be controlled and/or avoided. The electro-Leidenfrost effect presented in this letter will have to be fully characterised as it has been done for the electro-wetting effect. The electrical conductivity of water as well as the nature of the substrate on which it is standing should have important consequences on its efficiency.  Finally, we hope this communication will motivate theoretical and numerical studies concerning the asymmetric underneath shape put in evidence for Leidenfrost droplets without any applied voltage.

We dedicate this paper to the memory of Prof. Richard Kofman who initiated this study.

\end{document}